\documentclass[prb,twocolumn,amssym,nofootinbib,floatfix]{revtex4} 
\usepackage{graphicx,natbib}
\begin{document}
\title{Interference in interacting quantum dots with spin}
\author{Daniel Boese}
\email{dboese@tfp.physik.uni-karlsruhe.de}
\affiliation{Institut f\"ur Theoretische Festk\"orperphysik, Universit\"at
  Karlsruhe, D-76128 Karlsruhe, Germany}
\author{Walter Hofstetter}
\altaffiliation[New address: ]{Lyman Laboratory, Harvard University,
  Cambridge, MA 02138, USA}
\affiliation{Theoretische Physik III, Elektronische Korrelationen und
  Magnetismus, Universit\"at Augsburg, D-86135 Augsburg, Germany}
\author{Herbert Schoeller} 
\affiliation{Institut f\"ur Theoretische Physik A, RWTH Aachen, D-52056
  Aachen, Germany} 
\date{\today}
\begin{abstract}
We study spectral and transport properties of interacting quantum dots 
with spin. Two particular model systems are investigated: Lateral 
multilevel and two parallel quantum dots. In both cases 
different paths through the system can give rise to interference. 
We demonstrate that this strengthens the multilevel Kondo effect for 
which a simple two-stage mechanism is proposed. In parallel dots we 
show under which conditions the peak of an interference-induced
orbital Kondo effect can be split.
\end{abstract}
\maketitle

\section{Introduction}
Interference is one of the key phenomena of quantum physics. The prototype
experiment is the famous double slit experiment where interference between
two possible paths leads to an oscillatory pattern on the detection screen. In
those experiments the phase difference is of geo\-metrical nature, i.e. one of
the paths is longer. A phase difference can also be introduced due to an
enclosed magnetic flux. In mesoscopic physics such an experiment is referred
to as Aharonov-Bohm (AB) ring, where the current
through the AB ring shows oscillations as function of the magnetic field
threading the ring.  

\begin{figure}
\includegraphics[width=8cm]{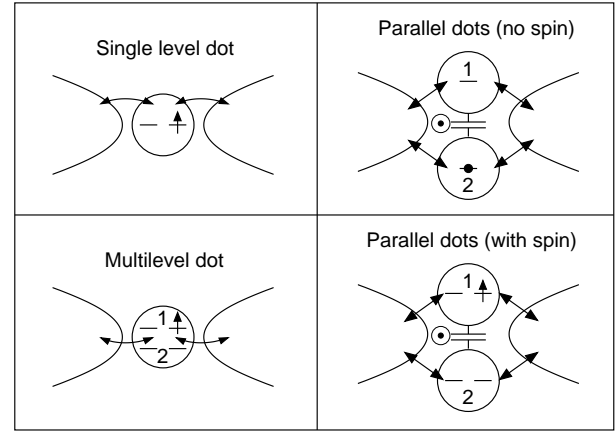}
        \caption{The four quantum dot setups of relevance to this
        work: Dot with one single, spin-degenerate level (top left),
        two parallel dots with one spinless level each, enclosing a
        flux (top right), a dot with two levels and spin (bottom left)
        and two parallel dots with one level with spin (bottom
        right). The paper is mainly concerned with the physics of the
        systems displayed in the bottom panels. 
}
        \label{fig:setup}
\end{figure}

\begin{figure}[t]
  \includegraphics[width=8cm]{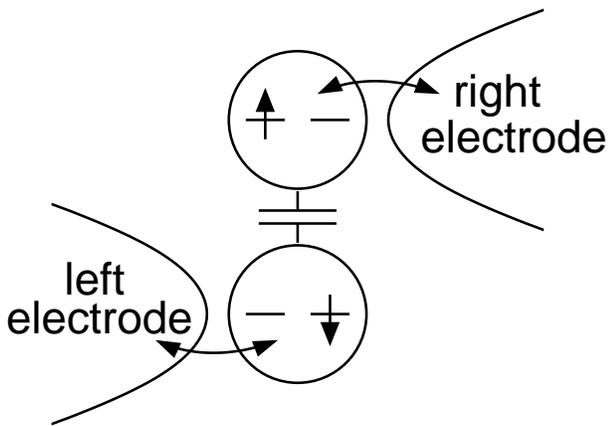}
        \caption{\label{fig:kondoschem}Destructive interference leads to a
          Kondo like situation. A 
          geometric (left/right) pseudospin is introduced. The quantum dots
          interact capacitively.}
\end{figure}

An AB ring can be used as an interferometer, where the object under
consideration is placed in one of the rings' arms, and the phase is tuned by
changing the object's parameters. In this way one can measure the transmission
phase of an interacting system, like a quantum
dot (QD),\cite{yacoby:95,schuster:97,wiel:00,ji:00,holleitner:01,levy:95,%
bruder:96,oreg:97}  
which in general (and especially when tuned to the Kondo regime) has a
complicated many-body ground state. In recent experiments quantum dots have
been put into both arms,\cite{holleitner:01} in some cases so close that a
strong capacitive Coulomb 
interaction between the two dots has been introduced (see 
Fig.~\ref{fig:setup} upper right for an illustration). The two paths are no
longer  
independent, but influence each other considerably. 
In a naive classical picture one could imagine that interaction would destroy
interference, as making use of one path effectively closes the
other. To answer this question the phase dependence of the current
needs to be studied, and it turns out that the current indeed can be
modulated. For completely equivalent paths ($\delta \epsilon = 0$ and
$T_1=T_2$) the system can be tuned opaque by setting $\phi=\pi$. In 
this case the Hamiltonian corresponds to a model of two capacitively coupled
QDs, each of which is coupled to a different reservoir (this can be seen from
the Hamiltonian in the form given in Eq.~\ref{eq:newH} and will be made
explicit in Sec.~\ref{sec:para}). Hence there is no way
for an electron to traverse from left to right (schematically shown in
Fig.~\ref{fig:kondoschem}). Note that such systems are of 
fundamental interest also because they can be viewed as artificial molecules
where e.g.~entangled states can be observed in transport and
noise.\cite{loss:00}   

The coherence of quantum mechanical states has recently become a topic of broad
interest, as it is fundamental to applications like quantum computing and to
many phenomena, such as the Kondo effect. In AB interferometers
coherence is essential as otherwise interference would not take
place. Therefore they constitute good test-grounds to study the gain and loss
of coherence in nanoscale devices, as was demonstrated by Buks and
coworkers\cite{buks:98} who demonstrated controlled dephasing by
intentionally introducing dephasing in one of the arms.

Single quantum dots can constitute interacting interferometers by
themselves. The capacitive Coulomb interaction between two dots is replaced
by the on-site interaction between different levels. 
The tunability of the phase
with magnetic fields, however, is lost, although some tunability using gates is
still present. Nevertheless it is instructive to study interference 
effects in single
quantum dots, since in general many dot levels participate in the
transport, see Fig.~\ref{fig:setup} bottom left. 
A prominent example is the occurrence of the Fano
effect\cite{goeres:00,hofstetter:01,bulka:01} with its characteristic lineshape,
which is due to interference between a resonant and a non-resonant transport channel.  
Moreover, it is often assumed that one level dominates the transport,
while the others are only very weakly coupled. We show that such a situation,
although not present in the beginning, can be created dynamically. 

In most quantum dots the levels are spin degenerate in the absence of a
magnetic field. The effect of this degeneracy is
manifold. As electrons with different spin can not interfere with each other
their role is contrary to interference. The difference is indeed drastic, as
on one side parallel QDs can be opaque due to destructive interference, while
on the other hand the spin in a single QD can form
a Kondo ground state leading to perfect
transparency.\cite{glazman:88,ng:88,wiel:00} 
Accounting for the spin degree of freedom is therefore a necessary step
towards more realistic models of QDs.  

In the course of this work we will show that the combination of 
interference and Kondo physics in multilevel QDs leads to a stronger
Kondo effect. However, this effect is caused by a new, effective 
level and thus resembles single level Kondo physics. 

Parallel QDs can be tuned to an interference--induced 
orbital Kondo effect by using the AB--phase.
We demonstrate that the corresponding Kondo peak is split only if both 
a magnetic field \emph{and} a level splitting are present. 

Interference can be described by a tunneling Hamiltonian with at least one
non-conserved index. Therefore the tunneling part takes the general form
$H_{\mathrm T} = \sum_{kr\sigma l n} T_{l n}^{kr} 
a_{kr\sigma n}^\dagger c_{\sigma n l} + {\mathrm h.c.}$.
The quantum number $l$ is present only in the QD Hamiltonian, it is
the analog of the paths. The index must not be conserved in tunneling, as
otherwise the 
electrons would not know of each other (as if they would be in different
reservoirs), ruling out any interference. $k$ denotes the
wavevectors and $n$ an additional conserved quantum number in reservoir $r$.
The conserved index $n$ can be due to
symmetries present in the leads and dot, such as a rotational symmetry in
some vertical quantum dots giving rise to an angular momentum quantum
number. As seen from the structure of the tunneling Hamiltonian, they play a
similar role as the spin and can cause and increase a Kondo
effect (orbital Kondo
effect).\cite{pohjola:00b,pohjola:01,wilhelm:00,wilhelm:01} In lateral quantum  
dots such 
symmetries are typically not present and we suppress those indices from now on.

Interference is also interesting from a technical and fundamental point of view.
The non-conservation of quantum numbers leads to non-vanishing off-diagonal
elements of the reduced density matrix of the local system, which describe the
coherence of states. Their presence explains why transport in first order,
which usually is referred to as sequential tunneling, can still be
coherent.\cite{koenig:01a} Moreover, non-equilibrium one-particle Green's 
functions are needed, even to describe the linear response regime.

The coupling to the leads can be so strong that perturbation
theory may not be sufficient anymore. For the Anderson model this is referred
to as the regime where Kondo correlations develop. Also for a simple model of two
spinless dot levels it has been shown that near destructive interference
the model can be mapped onto an effective Kondo model showing
strong-coupling behavior in a peculiar way. A phase transition of the type
RKKY vs Kondo tunable by a magnetic flux has been
predicted.\cite{boese:01d,boese:diss} 

In this work we study interference effects in strongly interacting quantum dot
systems with spin. In the next section we introduce and discuss the
model. In a qualitative discussion we summarize conclusions drawn from
a spinless model and generalize them to the present case. We then
focus on the Kondo effect multilevel QDs in Sec.~\ref{sec:multi} and
on the interference-induced orbital Kondo effect in parallel QDs in
Sec.~\ref{sec:para}. 
\section{\label{sec:model} Model}
We introduce
the following model Hamiltonian of two parallel, interacting QDs connected to
two electron reservoirs $r\in \{ \mathrm{R,L} \}$  via tunnel barriers, see
also Fig.~\ref{fig:setup} bottom right. Each
quantum dot (labeled $l \in \{1,2\}$) is modeled by an Anderson-type
Hamiltonian of a single spin-degenerate level
\begin{eqnarray}
\label{eq:hamiltonian}
H &=& \sum_{kr\sigma} \epsilon_{kr} a_{kr\sigma}^\dagger a_{kr\sigma} +
\sum_{l\sigma} \epsilon_{l} c_{l\sigma}^\dagger c_{l\sigma}  \\
&+& \sum_{(l \sigma) \neq (l'\sigma')}  U_{ll'} n_{l\sigma} n_{l'\sigma'} +
\sum_{krl  \sigma} \left( T_{l}^r a_{kr\sigma}^\dagger c_{l\sigma}
  +\mathrm{h.c.} 
  \right) . \nonumber 
\end{eqnarray}
The third term represents the Coulomb interaction, where $U_{ll}$ is of the
order of the intra-dot charging energy (in dot $l$), and $U_{12}$ reflects the
inter-dot
charging energy. To minimize the number of parameters involved we take
$U_{ll'}=U$, as they are similar in order of
magnitude.\cite{holleitner:01} We are interested in the case of strong 
interactions, i.e.\ when $U$ is the largest energy of the system, requiring an
explicit treatment. This allows to restrict the discussion on two charge
states, i.e.~$N \in \{ 0,1 \}$, and hence exchange terms may be
neglected.\footnote{This is not the case for $N > 1$, where
      interesting new physics can be observed \cite{hofstetter:02}.} 
The tunneling matrix elements $T_l^r$ are assumed to be
independent of spin and wavevector. If a magnetic flux is enclosed one can
either distribute the accumulated phase equally on the four $T_l^r$, or
equivalently attach the phase $\phi$ to one single element. We choose the
latter, i.e.\ we take $T_2^{\mathrm{L}} (\phi) = T_2^{\mathrm{L}} \exp (i
\phi)$,  and furthermore assume the matrix elements to be real and symmetric
with respect to left and right. Together with the density of states in the
leads $\rho_0$ (which is assumed to be independent of energy) we introduce the
coupling constants $\Gamma_{ll'}^r = 2 \pi T_l^r T_{l'}^{r,\ast} \rho_0$.  
The magnetic field shall be small enough, such that only the AB phase is
influenced, and Zeeman and orbital shifts can be neglected.

\begin{figure}
 \includegraphics[width=8cm]{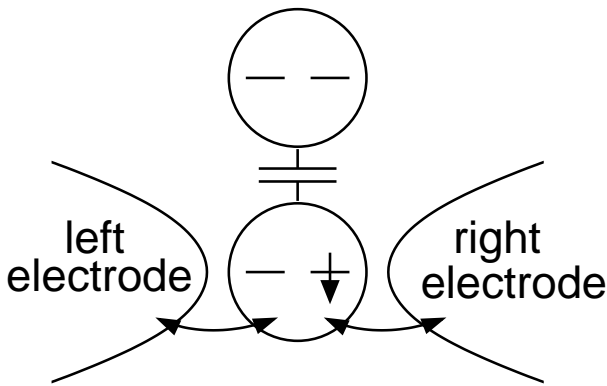}
        \caption{\label{fig:exmodel} For vanishing level spacing and phase,
          the QD can be mapped 
          onto a QD model as shown. Only one QD (the $f_{1\sigma}$
        level) is coupled to the leads. The 
          other one (the $f_{2\sigma}$ level) influences the transport
        only by electrostatic means. For 
          strong interactions the upper dot acts like a switch: When it is
          occupied the current is blocked, when it is empty, the lower dot
          behaves like a single dot. An exact solution of this model
        can be found in Ref.~\onlinecite{boese:diss}.}
\end{figure}

We introduce another set of dot states that simplifies the discussion later
on (see Fig.~\ref{fig:exmodel} for an illustration of the physical
meaning of these states)
 With $T_{1/2}$ being real (the $\phi$ dependence we take explicitly) and 
$\tau = \sqrt{T_1^2 + T_2^2}$ we can write 
\begin{equation}
\label{eq:newstates}
f_{1/2 \, \sigma} = \frac{ T_{1/2} c_{1\sigma} \pm T_{2/1} c_{2\sigma}}{\tau}.
\end{equation}
Together with the definition 
$\epsilon_{1/2} = \epsilon \pm \delta \epsilon/2$ 
this yields the new Hamiltonian
\begin{eqnarray}
\label{eq:newH}
H&=& \sum_{\sigma} \epsilon \left(n_{f_1 \sigma} +n_{f_2 \sigma}
   \right) -
\frac{\delta \epsilon \, T_1 T_2}{\tau^2} \left(f_{1\sigma}^\dagger
  f_{2\sigma} + f_{2\sigma}^\dagger f_{1\sigma} \right) \nonumber \\
 &+& \,\,\, U \!\!\!\!\!\!\! \sum_{(f_i \sigma) \neq (f_j \sigma')} \!\!\!n_{f_i
   \sigma}  n_{f_j \sigma'} + 
   \sum_{k \sigma} \left[ \tau a_{kR\sigma}^\dagger f_{1\sigma}
   \right.  \nonumber \\
&+& \left. a_{kL\sigma}^\dagger  \left( \frac{T_1^2+T_2^2 e^{i\phi}}{\tau}
   f_{1\sigma} 
   + \frac{T_1 T_2}{\tau} (1- e^{i \phi} ) f_{2\sigma} \right)
   \right. \nonumber \\
 &+& \left. \mathrm{h.c.} \right] +  \sum_{kr\sigma} \epsilon_{kr}
   a_{kr\sigma}^\dagger a_{kr\sigma} .
\end{eqnarray}
This makes clear that for $\delta \epsilon =0 $ the cases  $\phi=0$ and 
$\phi=\pi$ plus $T_1 = T_2$ are special and should be considered separately. 
Note that it is the DOS of the $f_{1\sigma}$ level that is relevant for the
transport. 

It is useful to compare the above Hamiltonian Eq.~(\ref{eq:hamiltonian}) to that
of a single, lateral, multilevel QD (see Fig.~\ref{fig:setup}b). In this case
the index $l$ labels the dot states
and the sum runs in general over many such states. Yet, for large level
spacing one may approximate the situation by taking only two states. A
generalization to many levels will be given in Section~\ref{sec:multi}. The
interaction parameters $U_{ll'}$ now corresponds to intra-dot interactions.
Taking them all equal is a standard assumption (constant interaction model).
Thus we see that, apart from the AB tunability, Eq.~(\ref{eq:hamiltonian}) also
describes multilevel, single QDs.

We note that this model goes beyond previous work. 
Inoshita et al.~\cite{inoshita:93} have considered 
only the case of vanishing AB phase, 
while the Coulomb interaction was treated approximately. 
In Ref.~\onlinecite{koenig:98b}, 
K\"onig and coworkers neglected interactions, phase dependencies and spin. In
a more recent work those were mostly accounted for,
their focus, however, was on the role of phase coherence in indecent
(i.e. non-interacting) arms of the AB ring.\cite{koenig:01a,koenig:01b}
Silvestrov and Imry\cite{silvestrov:00,silvestrov:01} investigated a
multilevel QD model (i.e.~no phase dependence), but
concentrated on the limit of one broad and one narrow level, utilizing
perturbative arguments. Their model of strongly and weakly coupled levels is
related to the Fano effect studied in Ref.~\onlinecite{hofstetter:01} and
\onlinecite{bulka:01} 
and measured by G\"ores and coworkers.\cite{goeres:00} In a previous work
of us,\cite{boese:01d} a more simple model, which neglects the spin, was
addressed. Models with spin but no dot-dot interaction have been studied in
Ref.~\onlinecite{izumida:97} and \onlinecite{loss:00}, while
in Ref.~\onlinecite{izumida:98}, which incorporates interaction, 
only special AB phases have been investigated, and Ref.~\onlinecite{ulloa:01}
is concerned with occupation numbers of the ground state.  

Our calculations are based on the numerical renormalization
group (NRG).\cite{wilson:75,costi:94,hofstetter:00}
\section{\label{sec:quali}Qualitative Discussion of General Properties}
We start with a discussion of multilevel dots with no phase, i.e.,
$\phi=0$. It is well known that QDs with a single level (the two-lead
Anderson model) display Kondo physics for temperatures below the
Kondo scale
\begin{equation}
\label{eq:tkdef}
T_K \sim \frac{\sqrt{U\Gamma}}{2}\exp \left( \frac{\pi \epsilon
(\epsilon+U)}{\Gamma U}\right) .
\end{equation}
The manifestation of this is an increased density of states at the
Fermi edge resulting in an increased conductance of the dot, which for
$T \rightarrow 0$ even may reach the unitary value of $2 e^2/h$. 
It is a priori not clear if and how this prevails when more orbitals
participate. 

The physics of two and more orbitals without spin has been addressed
before, and it was found that instead of Kondo physics a hybridization 
\begin{equation}
\label{eq:deltadef}
\Delta \sim \frac{\Gamma}{2\pi} \ln \frac{E_C}{\omega_c} 
\end{equation}
of the two levels is introduced.\cite{boese:01d,boese:diss} This scale
$\Delta$ is much larger than the exponentially small Kondo scale, and
it leads to a shoulder in the DOS of order $\Delta$ above the Fermi
edge. The weight of this shoulder is related to the level splitting
and vanishes for $\delta \epsilon \rightarrow 0$ and its width is
roughly half the width of the main excitation, i.e.~$\Gamma/2$.  

In order to understand what happens for two orbitals with spin we
perform a Schrieffer-Wolff transformation (see
App.~\ref{sec:schrieffer} for details), followed by a poor man's
scaling approach. In this transformation the hybridization is created and thus the
level splitting increases until it becomes of the same order as the
flow parameter $\omega_c$. Then the upper $f_{2\sigma}$ level is too
high in energy, decouples, and thus does not participate anymore. The scaling
proceeds with the renormalized single $f_{1\sigma}$ level. Hence we
have found a two-stage situation: First one level is pushed upwards
until it is out of reach, then in the second step the remaining,
renormalized level makes the Kondo effect alone. 

The picture is slightly different for the parallel QDs. The flux
enclosed leads to destructive interference and the current can even go
to zero. The energy scale $\Delta$ is modified by a factor
$(1+\exp[i\phi])/2$ and thus vanishes for $\phi=\pi$. In this case the
model can be mapped onto an effective Kondo model. When the spin is
included this is still the case and a more strong Kondo effect takes
place as will be discussed in Sec.~\ref{sec:parakondo}.

\section{\label{sec:multi} Multilevel quantum dots}
We now discuss multilevel QDs in detail. Quantum dots have in general many
levels that can participate in transport. In 
contrast to vertical QDs, the states in lateral QDs are labeled by a
non-conserved quantum number. Furthermore, a multilevel structure is also
relevant to 
other systems, like single atom contacts,\cite{kirchner:00} heavy fermion
compounds (e.g.~studied by photo-emission\cite{reinert:01}) or general
molecular electronics setup, where many channels can interfere. We
focus on the interesting regime of levels below the Fermi edge and low
temperatures. This is the regime of the Kondo effect, where
correlation effects dominate and the dot's spin is screened by the
electrons in the leads. For clarity we mention again that $\phi=0$ in this section.

\begin{figure}
  \includegraphics[width=8cm]{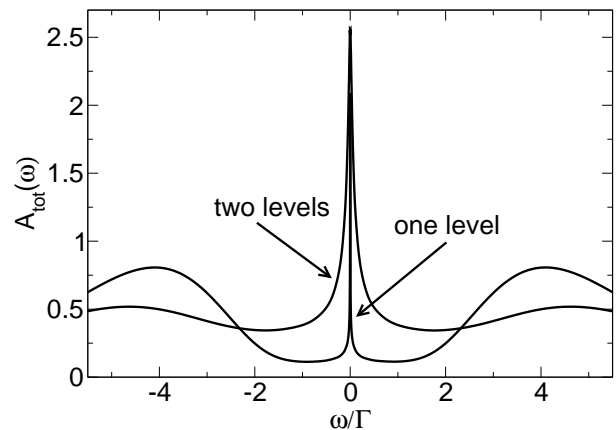}
        \caption{\label{fig:strongkondo} Effective density of states for the
          Kondo effect with one and two orbitals. The Kondo temperature
            increases strongly with the number of levels.  
            Parameters for the symmetric dot are in
            units of $\Gamma$: $2 \pi U=50$, $ \epsilon_1=\epsilon_2=-25 /2 \pi$,
          $2 \pi D=25$, $\phi=0$, $T=0$.}
\end{figure}

In a first step we look at the case of two degenerate levels in the
dot. In Fig.~\ref{fig:strongkondo} we show results for the 
total spectral density. There are
four possible states an electron can occupy in the dot, 
characterized by a spin index, which is conserved in tunneling, and an
orbital index, which is not conserved. 
As discussed before, this is
equivalent to one strongly coupled level and one decoupled one. Hence we see
single-level Kondo physics with greatly increased $T_K$. The big
increase of $T_K$ compared to the factor of $\sqrt{2}$ in the
tunneling matrix element can be easily understood from the definition
of $T_K$ which involves the coupling $\Gamma$ exponentially.

\begin{figure}
  \includegraphics[width=8cm]{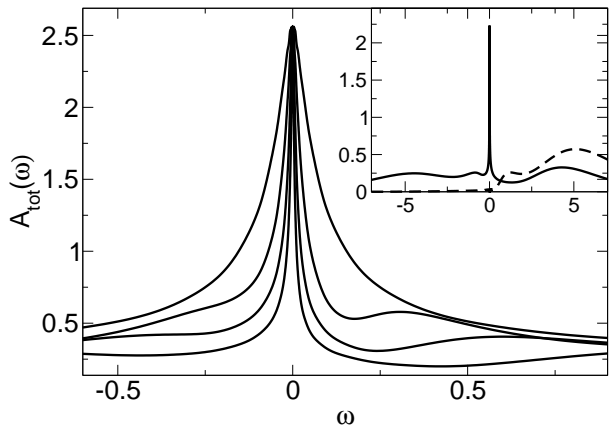}
        \caption{\label{fig:speckondo}Effective density of states for
          a multilevel Kondo dot with increasing level splitting. The lower
          level sits at $2 \pi \epsilon_1=-25$ and the upper level at
          $2 \pi \epsilon_2 = -25$, $-23.75$, $-22.5 $ and $-20$
          (outermost to innermost curve, everything in units of $\Gamma$). The
          inset shows the spectral 
          densities of the lower (solid) and upper level (dashed) for
          $2 \pi \epsilon_2 = -20$. Common parameters are  $2 \pi U=50$,
          $2 \pi D=25$, $\phi=0$, $T=0$. }  
\end{figure}

In the second step we allow the two orbitals to be different in
energy. One might speculate that this should lead to the appearance of
side or satellite Kondo peaks. However, in
Fig.~\ref{fig:speckondo} we demonstrate that single-level Kondo physics is 
effectively seen for split levels as well. 
With increasing splitting the Kondo
peak becomes narrower, signaling a decreasing $T_K$. At the same time the
shoulder discussed in the previous section becomes visible and progressively
moves to higher frequencies. This can be understood from the Schrieffer-Wolff
transformed Hamiltonian in the $f$-basis. Equation~(\ref{eq:swhamf}) shows
that only the $f_{1\sigma}$ level generates the Kondo resonance. In the scaling
language it can be thought of as a two-step 
process. First the tunnel-splitting is created from integrating out the very
high energies. This stops at an intermediate energy scale $\omega_c$, where 
diagonalization shifts one level above $\omega_c$. It can no longer contribute 
to scaling, while the other one -- the broad $f_{1\sigma}$ level -- 
stays in the window. The
scaling now gives the usual Kondo physics of a single, but modified level. It
should be noted that this reflects the strong coupling behavior of the
problem, i.e., all energy scales are important and contribute equally. In
the inset of Fig.~\ref{fig:speckondo} we show the partial spectral densities
of the upper and lower level which demonstrate that the lower level\footnote{For this
  level splitting the lower and the $f_{1\sigma}$ level have significant
  overlap.} 
alone produces the Kondo peak. The upper level is not occupied and does not
participate. 

\begin{figure}
  \includegraphics[width=8cm]{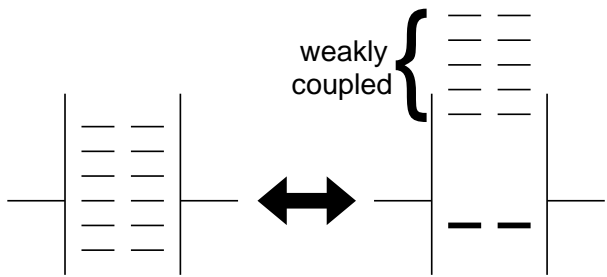}
        \caption{\label{fig:multilevel}Scheme of the effect of the
          renormalization group for a multilevel quantum dot: One broadened
          level remains while the others are 
          moved to higher energies and weaker coupling.} 
\end{figure}

This mechanism can be generalized to many ($N$) levels, where the role of the
$f_{1\sigma}$ level is played by the `sum' over or the superposition of all
levels. One level after the other 
is shifted to higher energies, and only one broad ($\sim N \Gamma$) level
remains, as sketched in Fig.~\ref{fig:multilevel}. This new, broad level alone
participates in the Kondo effect, which shows a strongly increased $T_K$,
making it much easier to observe. We suggest that this mechanism explains the
observed single-level Kondo physics in QDs. 

We conclude that even for many spin-degenerate levels (with
non-conserved orbital index) only one single Kondo peak is seen. The Kondo
temperature depends on the level splittings. The other excitations can
be traced back to shoulders as discussed in
Refs.~\onlinecite{boese:01d}, \onlinecite{boese:diss} and 
\onlinecite{inoshita:93}. In two
parallel QDs the level splitting is easily tunable, which  
allows to directly measure the change of $T_K$. 

\section{\label{sec:para}Parallel quantum dots}
In this section we study the physics of two
parallel, interacting quantum dots as previously introduced, which can be 
tuned by an AB phase. We focus on the special case $\phi=\pi$, which
corresponds to a Kondo--like situation. Note that this does
not necessarily require parallel QDs but can also be realized in
multilevel dots, when for instance one level is symmetric and the other
anti-symmetric. 

\subsection{\label{sec:parakondo} Interference-induced orbital Kondo effect} 
\begin{figure}
  \includegraphics[width=8cm]{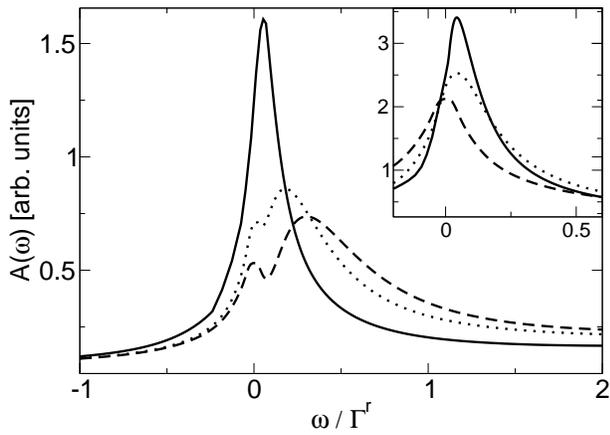}
        \caption{\label{fig:interkondo}Spectral density $A(\omega)$ of level
          $1$ (main panel) and 
          effective density of states (inset). The phase $\phi$ is changed
          from $0$ (dashed), over $\pi/2$ (dotted) to the value of the
          interference-induced orbital Kondo effect, $\phi=\pi$
          (solid). Parameters for the symmetric dots are in units of
          $\Gamma^r$: $U=50/\pi$, $\epsilon_1=\epsilon_2=-25 /\pi$,
          $D=25/\pi$, $T=0$. } 
\end{figure}

As mentioned before, the case $\phi=\pi$ 
corresponds to a model where one level couples only to the left and the other
one only to the right, as shown in Fig.~\ref{fig:kondoschem}. 
Evidently there are two conserved quantities: the spin and a
geometrical pseudo-spin (left/right). 
Introducing symmetric and antisymmetric combinations of the lead states 
$b_{ki\sigma} = a_{kR\sigma} - (-1)^i a_{kL\sigma}$, we can rewrite 
the tunneling part of the Hamiltonian as  
\begin{equation}
\label{eq:kondoHtun}
H_T = \sum_{ki\sigma} T_i b_{ki\sigma}^\dagger c_{i\sigma} + \mathrm{H.c.}\; .
\end{equation}
This has the form of an Anderson Hamiltonian with the two conserved
quantities discussed before. One therefore finds an enhanced Kondo effect 
for a low lying level at low temperatures. In other words, 
the state of complete destructive interference is a strong coupling state.
Such models have been studied for instance for multilevel
vertical quantum dots,\cite{pohjola:97} where the orbital momentum is
conserved in tunneling, 
or in double-layer QD system,\cite{wilhelm:00,wilhelm:01,pohjola:00b,pohjola:01}
where the index $i$ 
corresponds to the upper or lower plane. In such cases the Kondo temperature
is enhanced with respect to a pure spin Kondo model, as the second quantum
number -- the pseudospin -- can give rise to Kondo correlations alone. This is true also
in our case, where strong correlations can be expected even without
spin.  
In Fig.~\ref{fig:interkondo} we show the spectral
density corresponding to $c_{1\sigma}$. For zero phase a weak Kondo peak and
a second broader peak at higher frequencies are visible. The broad
peak (essentially the shoulder discussed before) moves to lower frequencies
when the phase is increased towards $\pi$ and merges with the Kondo resonance
for $\phi=\pi$. This strengthens the peak and thus enhances the Kondo
temperature $T_K$ as can be seen more clearly in the inset, where the
density of states of the $f_{1\sigma}$ level is shown. Note
that one of the special features of this Kondo effect is that the tunneling
matrix elements are tunable for each (pseudo)-spin, as the individual
levels can be controlled.

We remark that the Kondo effect discussed here is qualitatively different from
an orbital Kondo effect as discussed in Ref.~\onlinecite{pohjola:01} 
and also from two-channel Kondo physics.\cite{matveev:91,vladar:83,zawadowski:99,cox:98}

\subsection{\label{sec:peaksplit}Splitting the Kondo peak}
The ordinary Kondo effect in quantum dots can be destroyed by the application
of either a magnetic field that splits the level by the Zeeman energy
$\Delta_Z$ or by a
bias voltage introducing dephasing\cite{kaminski:99,kaminski:00,rosch:01} (where
the latter might under certain conditions open the door for two-channel Kondo
physics again\cite{coleman:01,rosch:01}). In our case the orbital Kondo effect 
can be destroyed by the analog of the Zeeman term which is the level splitting, by
different tunneling amplitudes (not accessible in ordinary QDs),  by a bias
voltage in the usual sense, and via a detuning of the phase,
i.e.~away from $\phi=\pi$. 

\begin{figure}
  \includegraphics[width=8cm]{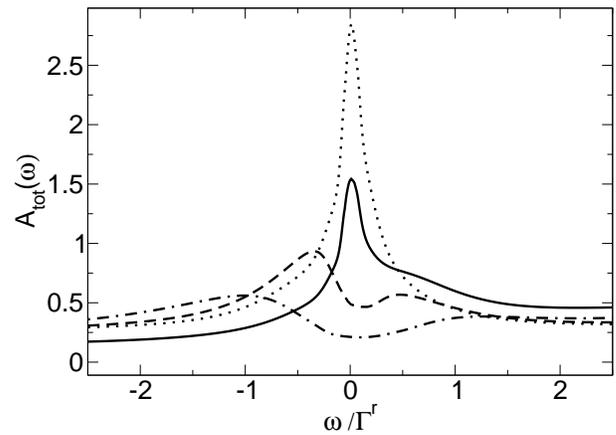}
        \caption{\label{fig:interkondo:split}
          Effective density of states at $\phi=\pi$ under the influence of
            Zeeman and level splitting. No splitting can be seen for the
            combination $\Delta_Z = 0.5$ and $\epsilon_2 = -5$ (dotted line)
            or for $\Delta_Z = 0 $ and $\epsilon_2=-5.5$ (solid line). If both
            splittings are introduced at the same time a splitting is seen for
            $\Delta_Z = 0.5 $ and $\epsilon_2=-5.5$ (dashed line) and
            $\Delta_Z = 1 $ and $\epsilon_2=-6$ (dot-dashed line).
            Parameters for the symmetric dots are in units of
          $\Gamma^r$: $U=10$, $\epsilon_1=\epsilon_2=-5$,
          $D=5$, $T=0$. } 
\end{figure}

An interesting question is whether a splitting of
the levels leads to a splitting of the Kondo peak, the development of
satellite peaks or if only a weakening and destruction of the Kondo peak is
observed. In Fig.~\ref{fig:interkondo:split} we find that a peak
splitting can 
only be observed if both, the Zeeman and the orbital level splitting, are
introduced. No side peaks appear if only one of them is present, which only
leads to a reduction of $T_K$. The suppression of side peaks has been
attributed to an enhanced dephasing rate, such as produced by
spinflip-cotunneling.\cite{meir:93,kaminski:99,kaminski:00} 

Note that this result also applies to other geometries like
double-layer QDs.\cite{pohjola:00b,pohjola:01,wilhelm:00,wilhelm:01} 

The detection of an inter\-ference-induced orbital Kondo effect is more
difficult than for the usual spin Kondo effect. Nevertheless it is possible by
probing the resonance by additional leads to the
dot.\cite{sun:01,lebanon:01,difrancesco} If 
the coupling is weak enough one can perform spectroscopic measurements on the
spectral densities in the 
individual dots. Another method is to measure the transport and noise
properties of a quantum point contact which is in the vicinity\cite{sprinzak:01}
of the double dot 
system. In contrast to the spin Kondo effect, the up and down pseudospins
correspond to charges in the upper or lower dot, which are much easier to
detect. The strong fluctuations in the Kondo regime will therefore influence the
transmission properties of the point contact allowing an indirect measurement
of the Kondo resonance, in a way which is not accessible for the usual spin Kondo
effect. The measurement of charge fluctuations thus provides a direct handle on
spin fluctuations.  

In real QD systems complete destructive interference, where the dots become
opaque, is not achieved experimentally. The reasons are the difficulty to
realize exactly equal QDs, as well 
as effects not captured in our model, such as more levels (at higher energy)
or processes that break the phase coherence of an otherwise coherent process
(less relevant at low temperatures). Yet, more than 50\% contrast is 
possible in today's experiments\cite{holleitner:01} and the effect
is therefore observable. 
\section{Conclusions} 
We studied coherence in two interacting quantum dot systems. 
First we investigated multilevel QDs with spin. We discussed the relevant
excitations and energy scales. The multilevel Kondo effect has been analyzed. We
demonstrated that single-level Kondo physics essentially prevails, and that
the corresponding Kondo temperature can be strongly enhanced. 
We have also investigated a very similar system, namely 
two single-level (but spin-degenerate) QDs in parallel.
Their behavior can be tuned by an enclosed magnetic flux. We showed that 
coherence persists when
the two dots interact with each other. 
In the case of destructive interference, the system
exhibits novel Kondo behavior (interference-induced orbital Kondo effect) that
is not due to the spin degree of freedom and 
allows to access Kondo correlations via charge fluctuations. Side peaks in the
density of states appear only if a Zeeman and a level splitting are
introduced together. 
\begin{acknowledgments}
We would like to thank S.~Kleff, J.~K\"onig, J.~Kroha, T.~Pohjola, A.~Rosch,
G.~Sch\"on, and D.~Vollhardt for useful discussions. This work was supported by the DFG
through Graduiertenkolleg ''Kollektive Ph\"anomene im Festk\"orper'' and the
CFN (D.B.), as well as through SFB 484 and a postdoctoral research grant (W.H.).  
\end{acknowledgments}
\appendix
\section{\label{sec:schrieffer} Schrieffer-Wolff transformation} 
We perform a unitary transformation on the Hamiltonian
Eq.~(\ref{eq:hamiltonian}) such that the un- and doubly-occupied states are
projected out
\begin{equation}
H'=e^S H e^{-S} = H_0 + \frac{1}{2} \left[ S, H_T \right] + \dots\; ,
\end{equation}
where $S$ has been chosen to fulfill $[S,H_0]=-H_T$. In our case this operator
is given by 
\begin{eqnarray}
S &=& \sum_{krs\sigma} T^{r,\ast}_{ks\sigma} \left( \frac{1- ( n_{\bar{s}
    \bar{\sigma}}  +  n_{\bar{s} \sigma} + n_{s
    \bar{\sigma}})}{\epsilon_{s\sigma}-\epsilon_{kr}} 
 \right. \nonumber \\ &+& \left. \frac{ n_{\bar{s}
    \bar{\sigma}}  +  n_{\bar{s} \sigma} + n_{s \bar{\sigma}}}
    {\epsilon_{s\sigma}+U -\epsilon_{kr}}\right) c^\dagger_{s\sigma}
    a_{kr\sigma} - \mathrm{h.c.} \; .
\end{eqnarray}
To avoid cluttering the notation we suppress the indices on the
tunneling matrix elements and local energies from now on, and take
$U\rightarrow \infty$. We introduce the two new coupling constants
\begin{eqnarray}
J_k &=& -   \frac{ |T|^2}{\epsilon-\epsilon_{k}} \\
\Delta_0 &=& \sum_{kr} J_k \; ,
\end{eqnarray} 
The new Hamiltonian is finally
given by 
\begin{eqnarray}
\label{eq:swham}
H &=& H_0 + \left[- \Delta_0 \sum_{ss'\sigma}  
  c^\dagger_{s\sigma}  c_{s'\sigma} + \sum_{krk'r's\sigma} \!\!\! J_k \,
  n_{s\sigma} 
  a^\dagger_{k'r'\sigma}  a_{kr\sigma} \right.\nonumber  \\ 
&+&  \sum_{krk'r's\sigma} \!\!\!J_k  \left( c^\dagger_{s\sigma}
 c_{\bar{s}\bar{\sigma}} a^\dagger_{k'r'\bar{\sigma}} a_{kr\sigma} +
 c^\dagger_{s\sigma} c_{\bar{s}\sigma} a^\dagger_{k'r'\sigma} a_{kr\sigma}
 \right. \nonumber \\ &+& \left. \left.
 c^\dagger_{s\sigma}  c_{s\bar{\sigma}}
 a^\dagger_{k'r'\bar{\sigma}}a_{kr\sigma}\right) \right]
\end{eqnarray}
Replacing the dot operators by the (anti)-symmetric combinations
$f_{1/2\sigma}$ we obtain
\begin{eqnarray}
\label{eq:swhamf}
H &=& H_0^{\mathrm{res}} + \frac{\epsilon_1 +\epsilon_2}{2} \left(
  f^\dagger_{1\sigma} 
  f_{1\sigma} + f^\dagger_{2\sigma} f_{2\sigma} \right) 
   \\ &+& 
  \sum_\sigma \delta 
  \epsilon \left(f^\dagger_{1\sigma} f_{2\sigma} +\mathrm{h.c.}\right)
  -  
  \Delta_0 \sum_{\sigma}  f^\dagger_{1\sigma}  f_{1\sigma}  \nonumber \\
&+& \sum_{krk'r'\sigma} J_k \left(a^\dagger_{kr\sigma} a_{k'r'\sigma}
  f^\dagger_{1\sigma} f_{1\sigma} + a^\dagger_{kr\bar{\sigma}} a_{k'r'\sigma}
  f^\dagger_{1\sigma} f_{1\bar{\sigma}} \right) \; . \nonumber 
\end{eqnarray}

\end{document}